\documentclass[11pt,twoside]{article}

\usepackage{asp2014}

\aspSuppressVolSlug
\resetcounters

\usepackage{media9}

\begin{document}

\title{ESASky v. 2.0: all the skies in your browser}

\author{Bruno Mer\'{\i}n$^1$, Fabrizio Giordano$^2$, Henrik Norman$^3$, Elena Racero$^2$,
Deborah Baines$^4$, Jes\'us Salgado$^4$, Bel\'en L\'opez Mart\'{\i}$^5$, Sara Alberola$^1$, 
Marcos L\'opez Caniego$^6$, Ivan Valtchanov$^6$, Guido de Marchi$^7$
\& Christophe Arviset$^1$
\\
\affil{$^1$ ESAC Science Data Centre, ESA-ESAC, Spain \email{Bruno.Merin@esa.int}}
\affil{$^2$ ESAC Science Data Centre, SERCO for ESA-ESAC, Spain}
\affil{$^3$ Winter Way AB for ESA-ESAC, Sweden}
\affil{$^4$ ESAC Science Data Centre, QUASAR for ESA-ESAC, Spain}
\affil{$^5$ RHEA for ESA-ESAC, Spain}
\affil{$^6$ European Space Astronomy Centre (ESAC), ESA, Spain}
\affil{$^7$ ESAC Science Data Centre, ESA-ESTEC, The Netherlands}}

\paperauthor{Bruno Mer\'{\i}n$^1$}{bruno.merin@esa.int}{0000-0002-8555-3012}{European Space Agency}{Science Operations Department}{Villanueva de la Ca\~nada}{Madrid}{28692}{Spain}
\paperauthor{Fabrizio Giordano}{Fabrizio.Giordano@esa.int}{}{European Space Agency}{Science Operations Department}{Villanueva de la Ca\~nada}{Madrid}{28692}{Spain}
\paperauthor{Henrik Norman}{}{}{Winter Way AB}{}{Uppsala}{}{}{Sweden}
\paperauthor{Elena Racero}{Elena.Racero@esa.int}{}{European Space Agency}{Science Operations Department}{Villanueva de la Ca\~ada}{Madrid}{28692}{Spain}
\paperauthor{Deborah Baines}{Deborah.Baines@esa.int}{}{European Space Agency}{Science Operations Department}{Villanueva de la Ca\~ada}{Madrid}{28692}{Spain}
\paperauthor{Jes\'us Salgado}{Jesus.Salgado@esa.int}{}{European Space Agency}{Science Operations Department}{Villanueva de la Ca\~ada}{Madrid}{28692}{Spain}
\paperauthor{Bel\'en L\'opez Mart\'{\i}}{}{}{}{}{}{Madrid}{}{Spain}
\paperauthor{Sara Alberola}{}{}{European Space Agency}{Science Operations Department}{Madrid}{}{}{Spain}
\paperauthor{Marcos L\'opez Caniego}{mlopez@sciops.esa.int}{}{European Space Agency}{Science Operations Department}{Villanueva de la Ca\~nada}{Madrid}{28692}{Spain}
\paperauthor{Ivan Valtchanov}{ivaltchanov@sciops.esa.int}{}{European Space Agency}{Science Operations Department}{Villanueva de la Ca\~nada}{Madrid}{28692}{Spain}
\paperauthor{Guido de Marchi}{guido.de.marchi@esa.int}{}{European Space Agency}{Science Operations Department}{Villanueva de la Ca\~nada}{Madrid}{28692}{Spain}
\paperauthor{Christophe Arviset}{Christophe.Arviset@esa.int}{}{European Space Agency}{Science Operations Department}{Villanueva de la Ca\~nada}{Madrid}{28692}{Spain}

\begin{abstract}
With the goal of simplifying the access to science data to scientists and citizens, ESA recently released ESASky (http://sky.esa.int), a new open-science easy-to-use portal with the science-ready Astronomy data from ESA and other major data providers. In this presentation, we announced version 2.0 of the application, which includes access to all science-ready images, catalogues and spectra, a feature to help planning of future JWST observations, the possibility to search for data of all (targeted and serendipitously observed) Solar System Objects in Astronomy images, a first support to mobile devices and several other smaller usability features. 
We also discussed the future evolution of the portal and the lessons learnt from the 1+ year of operations from the point of view of access, visualization and manipulation of big datasets (all sky maps, also called HiPS) and large catalogues (like e.g. the Gaia DR1 catalogues or the Hubble Source Catalogue) and the design and validation principles for the development of friendly GUIs for thin layer web clients aimed at scientists. 
\end{abstract}

\vspace{-0.5cm}


\section{Introduction: ESASky concept}

With the goal of maximizing the exploitation of science data obtained with telescopes and infrastructures made possible with public funding, ESA has developed ESASky\footnote{\url{http://sky.esa.int}} at ESAC, Madrid, Spain, by the ESAC Science Data Centre (ESDC, \citeauthor{Arviset2015} 2015)\footnote{\url{http://www.cosmos.esa.int/web/esdc}}, with collaboration and support from many ESA Mission science and technical experts, and from the Centre de Donn\'ees Astronomiques de Strasbourg (CDS)\footnote{\url{http://cdsweb.u-strasbg.fr}}. 

\begin{figure}
\includegraphics[width=\textwidth]{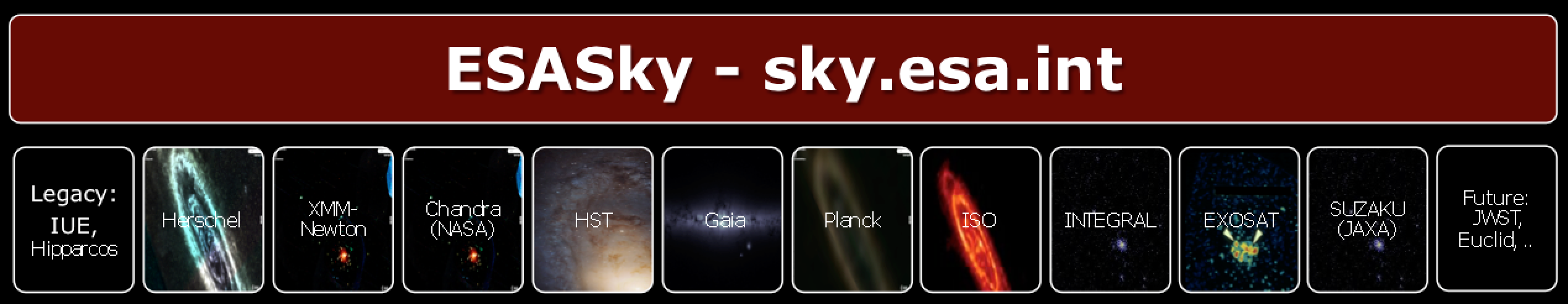}
\label{fig:ESASky_Design}
\caption{Design of the ESASky application. A web application sits in between the user and the mission archives, selecting fully calibrated high-level science-ready data files from the archives and serving them to the user in a simplified way.}
\end{figure}

We have designed a web application that stands in front of all the mission archives and presents and serves to the user selected high-level (i.e. fully processed and calibrated) data FITS files from those mission archives in a simplified way to the users (see Figs. \ref{fig:ESASky_Design} and 2). To allow simple visualization, the application projects all science data from the individual archives in the {\it same view} of the sky to facilitate comparison while preserving the original resolution. 

\section{Data holdings}
\label{data_holdings}

This section describes the astronomical datasets available for download and/or visualization. ESA has the purpose of having in ESASky always the latest and highest quality public science-ready data from all the ESA Astronomy missions. The current version gives users access to the following fully-reduced science-ready public data products:

\begin{itemize}
\item Access to science-ready images from XMM-Newton, SUZAKU, Chandra, HST, ISO and Herschel\footnote{Find all info about the ESA and ESA collaborative missions at \url{https://www.cosmos.esa.int/our-missions} and about NASA's Chandra mission at \url{http://chandra.harvard.edu/}. The description of the science-ready data products offered is at \url{https://www.cosmos.esa.int/web/esdc/esasky-help}}.
\item Access to spectroscopic science ready-products from XMM-Newton, Chandra, IUE, HST, ISO and Herschel.
\item Visualization and download of the following catalogues: Gaia DR1\footnote{Find all reference publications and contents info of all catalogues in the tool at the online documentation: \url{https://www.cosmos.esa.int/web/esdc/esasky-help}}. and Gaia TGAS, Hipparcos-2, Tycho-2, INTEGRAL IBIS/ISGRI Soft Gamma-Ray Source Catalog, XMM-Newton catalogues: 3XMM DR7, XMMSL1.6, XMM-SUSS2.1, Hubble Source Catalog v2.1, Herschel mission point source catalogues for PACS and SPIRE, Planck mission catalogues: Planck Catalogue of Compact Sources (PCCS2), Planck Sunyaev-Zeldovich 2 (SZ2), Planck Galactic Cold Cores Catalogue (PGCC2) and Chandra Source Catalog v1.1.
\item Progressive all-sky maps (HiPS) from INTEGRAL IBIS, EXOSAT LE,  XMM-Newton EPIC and OM, HST ACS, WFPC2, WFC3, NICMOS, FOC, WFPC, ISO ISOCAM, AKARI FIS, Herschel PACS and SPIRE, Planck HFI and LFI, Haslam 408 MHz and other selected surveys from Gamma-rays to radio wavelengths from the large HiPS collection curated by the Centre de Donn\'ees Astronomiques de Strasbourg (CDS)\footnote{\url{http://aladin.u-strasbg.fr/java/nph-aladin.pl?frame=aladinHpxList}}. See \citet{Fernique2015} for a detailed description of the HiPS format.
\end{itemize}


\articlefigure[width=0.9\textwidth]{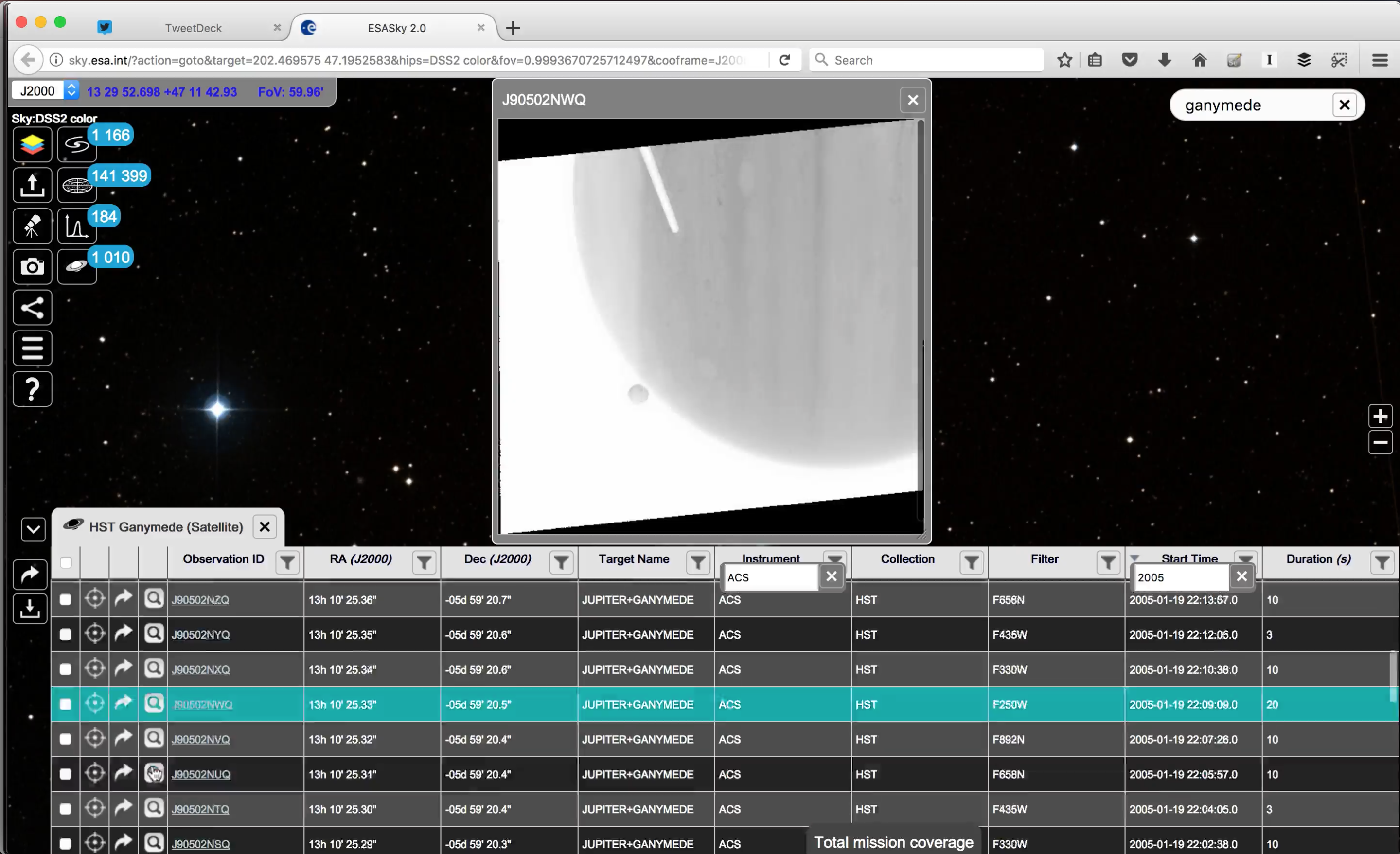}{ex_fig1}{Snapshot of ESASky v.2.0: Visualizing Ganymede raising behind Jupiter as observed by HST/ACS in a few clicks}

All the data products served by ESASky have been identified by the mission specialists as the best single FITS files per observation that they would give for doing research to a scientist who is not a specialist in the mission.

\section{Experience from the first 1.5 years of operations}

Since the release of ESASky 1.0 in May 2016, the application has received over 30,000 individual users from the whole world, with peaks of usage at announcements or at astronomical events, such as e.g. the search for electromagnetic counterparts to the Neutron Star merger event GW170817 \citep{Abbott2017}.

The application currently filters large number of sources from catalogues searches and only shows the brightest 2000 in the field of view. This is an incremental approach to showing large catalogues source densities in large areas of the sky.

Users are encouraged to request features, vote for the most interesting requests and have a conversation about the future of the tool in our user forum\footnote{\url{http://esasky.userecho.com}}. These requests are evaluated by the ESDC development team to identify the features with the highest user-value over development-cost ratio, which are then developed first.

\section{New features in ESASky v2.0}

The following short videos demo the new features in version 2 of the application:

\begin{itemize}
\item How to download science-ready images and spectra for one target in ESASky\footnote{\url{https://youtu.be/4wCdvPGI_z0}}
\item How to visualise the sky in different wavelengths using ESASky\footnote{\url{https://youtu.be/zkJkhSDr0nQ}}
\item How to visualise a target list in ESASky\footnote{\url{https://youtu.be/M-aJn5TTd50}}
\item How to visualise the full sky coverage of a mission using ESASky\footnote{\url{https://youtu.be/FXfR9fNkBJg}}
\item How to plan future JWST observations with ESASky\footnote{\url{https://youtu.be/n2iW1HfVgVE}}
\item How to search for Solar System Objects in ESASky\footnote{\url{https://youtu.be/ct53j7EuokA}}
\item How to query ESASky with a python script using astropy/astroquery: cone search\footnote{\url{https://youtu.be/w_kO5PIWDxM} and python astropy module at \url{http://astroquery.readthedocs.io/en/latest/esasky/esasky.html}}
\item How to query ESASky with a python script using astropy/astroquery: single object\footnote{\url{https://youtu.be/10WeEcVQlWo}}
\end{itemize}

\section{Acknowledgments}

Please acknowledge publications done with the tool\footnote{\url{https://www.cosmos.esa.int/web/esdc/esasky-credits}}. This URL lists all the people who have contributed to its creation.

\end{document}